\documentclass[12pt]{article}
\def\psl{\hbox{/\kern-.5800em$p$}}
\newcommand{\Zparity}{\ensuremath{\mathbb{Z}_2}}

\def\gappeq{\mathrel{\rlap {\raise.5ex\hbox{$>$}}
{\lower.5ex\hbox{$\sim$}}}}
\def\lappeq{\mathrel{\rlap{\raise.5ex\hbox{$<$}}
{\lower.5ex\hbox{$\sim$}}}}
\newcommand{\im}{i}

\usepackage{caption2}

\usepackage{amsmath,amssymb}
\DeclareMathOperator{\Res}{Res}
\usepackage{axodraw}

\begin{document}
\oddsidemargin -0.4cm
\evensidemargin -0.8cm

\pagestyle{empty}
\begin{flushright}
UAB-FT-527\\
May 2002
\end{flushright}
\vspace*{5mm}

\begin{center}

{\Large\bf  Fayet-Iliopoulos terms in 5d theories }
\vspace*{5mm}
{\Large\bf and their phenomenological  implications }\\
\vspace{2.0cm}

{\large Daniel Mart{\'\i} and  Alex Pomarol}\\

\vspace{.6cm}
{\it IFAE, Universitat Aut{\`o}noma de Barcelona, 
08193 Bellaterra (Barcelona), Spain}
\vspace{.4cm}
\end{center}

\vspace{1cm}
\begin{abstract}
Supersymmetric theories in 5d orbifold
can have Fayet-Iliopoulos (FI) terms 
on the boundaries.
These terms  induce a 5d mass 
for the bulk matter fields.
We analyze this effect using $N=1$ superfields 
and show how it can be  understood as a mixing between
the U(1)  gauge boson and the graviphoton. 
We also study the implications of the FI-terms
in the mass spectrum of theories with 
Scherk-Schwarz supersymmetry breaking.
We present a new 5d supersymmetric model that has the 
massless spectrum of the Standard Model and predicts
 a Higgs boson  with a mass around the present experimental limit.
\end{abstract}

\vfill
\eject
\pagestyle{empty}
\setcounter{page}{1}
\setcounter{footnote}{0}
\pagestyle{plain}


\section{Introduction}

The presence of compact extra dimensions 
allows for new mechanisms of symmetry breaking.
One of the  simplest idea consists in breaking 
symmetries by boundary conditions \cite{ss}.
Recently there has been a renewed interest
to apply this idea 
to obtain 
the Standard Model (SM)
from 
 grand unified  theories or    supersymmetric theories in
higher dimensions.

Here we want to consider the effect 
of Fayet-Iliopoulos (FI) terms
in 5d models. 
It was  shown in Ref.~\cite{agw}
that FI-terms can be present on the boundaries of  a 5d orbifold
leading to 5d odd masses for the hypermultiplets.
We will study this effect in more detail 
using
 $N=1$ superfields \cite{agw,mp,heb}
and   introducing the radion superfield $T$.
This allows to make contact with supergravity and show
that the effect of the FI-term  can be understood as a mixing between
the graviphoton and U(1) gauge boson. 
We will also show how this mixing can be generated at the one-loop level.

We also want to analyze the modifications 
in the  mass spectrum of the hypermultiplet
due to the existence of a 5d odd mass (generated by the FI-term).
We  will  consider the case  where supersymmetry is broken by the
Scherk-Schwarz mechanism.
The presence of 5d odd masses
 leads to interesting variations on previous  models \cite{ab,pq,bhn}.
These effects  were first discussed in Ref.~\cite{fi1,bhn2} where it was shown
 that the massless spectrum can be modified becoming sensitive to
the ultraviolet cutoff.
We will present a new SS model where the effect of the 5d mass
is used to explain the large ratio $m_t/m_b$.
The massive spectrum is completely determined as a function of the radius
$R$.
Although large uncertainties do not allow to predict the scale
of electroweak breaking, we find that the model predicts a Higgs boson
of  mass close to the present experimental bound.

Some of the issues studied here were also 
discussed recently in
Refs.~\cite{fi2,fi3}.

\section{FI-terms in 5d supersymmetric theories}

FI-terms are tadpoles of the auxiliary fields of an abelian
gauge multiplet. 
In 4d supersymmetric theories 
these terms are very important since 
their presence 
implies a breaking of either supersymmetry or the gauge symmetry.
In 4d they  
arise from the operator  $ \int d^4\theta\, V$ where $V$ is the 
gauge superfield.
This operator is invariant under global supersymmetry and 
gauge symmetry.
Nevertheless,   if supersymmetry is  local (supergravity)
such a term can only be  gauge invariant if the  gauge
symmetry is an $R$-symmetry.

Here we want to analyze the role of FI-terms in 5d theories.
For that purpose it is useful to 
work with
$N=1$ superfields \cite{agw,mp,heb}.
Following Ref.~\cite{mp}, 
we can write the 5d supersymmetric abelian theory as 
\begin{equation}
    S_5 = \int d^{5}\! x\, 
\left[ \frac{1}{4g^2_5} \int d^2\theta\, T
        W^{\alpha}W_{\alpha} + {\rm h.c.}+ \frac{2}{g^2_5} \int
         d^4\theta\: \frac{1}{(T +
          T^{\dag})} \left( \partial_5 V - \frac{1}{\sqrt{2}} (\chi +
            \chi^{\dag})\right)^2 \right]\, ,
\label{GaugeAb}
\end{equation}
where $V$ is a 
$N=1$ 
vector supermultiplet and ${ \chi}$
a chiral
supermultiplet: 
\begin{align}
    V & = -\theta \sigma^{\mu} \bar{\theta} A_{\mu} -
    i\bar{\theta}^2\theta\lambda_{1} +
    i\theta^2\bar{\theta}\bar{\lambda}_{1} +
    \frac{1}{2} \bar{\theta}^2\theta^2 D\, ,\nonumber \\
    \chi &= \frac{1}{\sqrt{2}} \left(\Sigma + i A_{5}\right) +
    \sqrt{2} \theta \lambda_{2} + \theta^{2} F_\chi\, .
\end{align}
$V$ is given in the Wess-Zumino gauge.
Both $V$ and $\chi$ form the 5d gauge superfield.
Under a gauge transformation, the superfields transform as
\begin{align}
V&\rightarrow V+\Lambda+\Lambda^{\dag}\, ,\nonumber\\
\chi&\rightarrow \chi+\sqrt{2}\partial_5\Lambda\, ,
\label{gauget}
\end{align}
where $\Lambda$ is an arbitrary chiral field. 
The chiral superfield $T$ is the radion multiplet
\begin{equation}
 T=  R + i B_{5} + \theta \Psi^5_R + \theta^{2} F_T\, .
\label{radion}
\end{equation}
It is
useful to introduce $T$  to define the 5d gravitational background
on which the superfields propagate.
It also helps to make the connection with supergravity.
Here we are considering that the extra dimension is  flat and compact:
\begin{equation}
    ds^2 = \eta_{\mu\nu} dx^\mu dx^\nu + R^2 dy^2 \, .
\label{flat}
\end{equation}
We leave  the case of warped extra dimensions for  further publication.

If the theory is compactified in a circle $S^1$, $0\leq y\leq 2\pi$,
the following  term 
can also be present in the 5d lagrangian
\begin{equation}
\int d^2\theta\, \chi +{\rm h.c.}\, .
\label{fis1}
\end{equation}
This term is a tadpole for the auxiliary field  $F_\chi$ 
and therefore corresponds to a FI-term in 5d.
Notice that if this term is zero in the original theory, it cannot
be generated by radiative corrections 
since it is protected by supersymmetry.

If the extra dimension
is compactified in the  orbifold $S^1/\Zparity$ , $0\leq y\leq \pi$,
the term (\ref{fis1}) is not allowed by gauge invariance.
In this  case, however, another tadpole term 
can be present in the 5d lagrangian:
\begin{equation}
-\int d^4\theta\,  \xi\,\epsilon(y) \Big[\partial_5 V-\frac{1}{\sqrt{2}}
\left(\chi+\chi^\dagger\right)
\Big]\, ,
\label{fio}
\end{equation}
where $\epsilon(y)$ is a step function, $\epsilon(y)=1 (-1)$   for positive
(negative) $y$.
Notice that 
 the presence of  $\chi$  is necessary to make
Eq.~(\ref{fio}) gauge invariant under Eq.~(\ref{gauget}).
Integrating by parts, we can write the tadpole of $V$
in Eq.~(\ref{fio}) as
\begin{equation}
\int d^4\theta\, 2\xi\, \Big[\delta(y)-\delta(y-\pi)\Big]\,  V\, .
\label{fit}
\end{equation}
If the term  Eq.~(\ref{fit})
is not present in the original theory, it can be generated
at the one-loop level. 
Let us write the induced FI-term as
\begin{equation}
\int d^4\theta\, 2\xi_{in}(y)
\,  V\, .
\label{fi}
\end{equation}
We will see that Eq.~(\ref{fi}) can be generated by either boundary or bulk
fields.  
They  give a divergent contribution  that  renormalizes
the operator of  Eq.~(\ref{fit}).

\noindent {\bf Boundary fields:}
Given a  field $Q$
of charge $q_Q$   localized on
the boundary at $y=y^*$
\begin{equation}
S_5=\int d^{5}\! x\, 
\int d^4\mspace{-2mu}\theta\, Q^{\dag} e^{q_QV} Q\, \delta(y-y^*) \, ,
\label{bulkbound}
\end{equation}
 we have (see diagram 
of Fig.~1) the contribution
 \begin{equation}
\xi_{in}(y)
=q_Q\frac{\Lambda^2}{16\pi^2}\delta(y-y^*)\, .
\label{fiob}
\end{equation}
This contribution is ultraviolet divergent and depends quadratically
on the cutoff $\Lambda$.

\begin{figure}[htbp]
    \begin{minipage}[t]{0.5\linewidth}
        \centering
        \begin{picture}(100,40)(0,0)
            \Photon(0,20)(50,20){3}{7}
            \CArc(70,20)(20,0,360)
            \Vertex(50,20){2}
            \Text(-3,22)[r]{$V$}
            \Text(94,20)[l]{$Q$}
            \Text(48,28)[br]{$\delta(y-y^*)$}
        \end{picture}
        \caption{}
 \label{fig:tadpoleone}
    \end{minipage}
    \begin{minipage}[t]{0.5\linewidth}
        \centering
        \begin{picture}(100,40)(0,0)
            \Photon(0,20)(50,20){3}{7}
            \CArc(70,20)(20,0,360)
            \Vertex(50,20){2}
            \Text(-3,22)[r]{$V$}
            \Text(94,20)[l]{$\Phi,\Phi^{c}$}
                  \end{picture}
        \caption{} 
\label{fig:tadpoletwo}
    \end{minipage}
\end{figure}

\noindent{\bf Bulk fields:}
We will  consider a charged hypermultiplet \cite{mp}  
\begin{equation}
    S_5 = \int d^{5}\! x\, \left\{ \int d^4\mspace{-2mu}\theta\: \frac{1}{2}
    (T + T^{\dag} ) \,\left(\Phi^{\dag} e^{q_\Phi V} \Phi + 
\Phi^{c} e^{-q_\Phi V} \Phi^{c\, \dag}\right) + \int
    d^2\mspace{-2mu} \theta\, \Phi^{c} \Bigl[
    \partial_{5}+\frac{q_\Phi}{\sqrt{2}}\chi\Bigr] \Phi +  
\text{h.c.} \right\}\, ,
\label{eq:hyperS}
\end{equation}
with two 
possible boundary conditions for  the superfields:
\footnote{In the  $S^1/(\Zparity\times\Zparity^\prime)$ 
orbifold \cite{bhn} the case (a) corresponds
to the parity assignment   
$++$ and $--$ to $\Phi$ and $\Phi^c$  respectively, while case
 (b)  corresponds
to the parity assignment
$+-$ and $-+$.}

\noindent a) $\Phi$ is even 
 with respect to the two boundaries at $y=0$ and $y=\pi$,
while   $\Phi^c$ is  odd. 
We will denote them
 by $\Phi^{++}$ and $\Phi^{c\, --}$.

\noindent b)  $\Phi$ is even (odd)
 with respect to the boundary at $y=0$ ($y=\pi $),
while   $\Phi^c$ is  odd (even).
We will denote them
 by $\Phi^{+-}$ and $\Phi^{c\, -+}$.

\noindent 
The cases (a) and (b) correspond to impose respectively
periodic and antiperiodic
boundary conditions  for the superfields
under  a $2\pi $ rotation of the extra dimension.
Let us now calculate 
the contribution from $\Phi$ and $\Phi^c$  (Fig.~2) 
to the operator of Eq.~(\ref{fi}).
This is given by
\begin{equation}
\xi_{in}(y)
=q_\Phi R\int \frac{d^4p}{(2\pi)^4} 
\left[G^{\phi}_p(y,y)-G^{\phi^{c}}_p(y,y)\right]\, ,
\label{xii}
\end{equation}
where $G^{\phi,\phi^c}_p(y,y)$ are the scalar propagators of the
hypermultiplet.
It is instructive (to understand the origin of  $\xi_{in}(y)$)
to calculate these propagators
starting from an infinite extra dimension. 
In this case the propagator  of a 5d scalar with 4d Euclidean momentum $p$
is given by 
\footnote{Here we include $R$  to keep $y$ dimensionless.}
\begin{equation}
G_p(y,y^{\prime})=\frac{e^{-p|y-y^{\prime}|R}}{2p}\, .
\end{equation}
Compactifying in a circle means identifying $y\leftrightarrow y+2\pi n$
where $n\in Z$.
Therefore in the tadpole contribution (Fig.~2)
the fields  can  propagate  from $y$ to $y+2\pi n $.
If we now 
orbifold the space,   we  identify $y\leftrightarrow -y$.
The fields can   propagate from  $y$ to $-y+2\pi n $.
Therefore the propagator  in Eq.~(\ref{xii})  reads, for the case (a),
\begin{equation}
G^{\phi^{++},\phi^{c\, --}}_p(y,y)
=\sum_{n=-\infty}^\infty\Big[G_p(y,y+2\pi n )\pm
G_p(y,-y+2\pi n )\Big]\, .
\label{propa}
\end{equation}
Only the second term in Eq.~(\ref{propa})  gives a nonzero contribution
to Eq.~(\ref{xii}).
This is given by
\begin{equation}
\xi_{in}(y)
=q_\Phi R
\sum_{n=-\infty}^\infty
\int \frac{d^4p}
{(2\pi)^4}
\frac{e^{-2p|y-\pi n |R}}{p}\, .
\label{fib}
\end{equation}
This contribution is convergent away from 
the fixed points at $y=n\pi $ 
due to the exponential suppression.
At the fixed points $y=n\pi $, however, the integral diverges. 
The divergence depends on how we regularize the boundaries.
In the   thin-boundary limit, we get for the divergent
contribution (restricting now to the interval [0,$\pi$]):
\begin{equation}
\xi_{in}(y)=q_\Phi\frac{\Lambda^2 }{32\pi^2}
\Big[ \delta(y)+\delta(y-\pi )\Big]
+q_\Phi\frac{\ln(\Lambda/\mu) }{64\pi^2 R^2}
\Big[ \delta^{\prime\prime}
(y)+\delta^{\prime\prime}
(y-\pi )\Big]
\, ,
\label{fiH}
\end{equation}
where $\mu$ is an infrared cutoff.
For the case (b), where the fields are antiperiodic under
a $2\pi $ rotation,
a  factor $(-1)^n$ must multiply  the  propagator 
\begin{equation}
G^{\phi^{+-},\phi^{c\, -+}}_p(y,y)=\sum_{n=-\infty}^\infty(-1)^n
\Big[G_p(y,y+2\pi n )\pm
G_p(y,-y+2\pi n )\Big]\, ,
\label{propb}
\end{equation}
and, consequently, also in  the induced FI-term. Restricted to  the interval
[0,$\pi $], we have 
\begin{equation}
\xi_{in}(y)=q_\Phi\frac{\Lambda^2 }{32\pi^2}
\Big[ \delta(y)-\delta(y-\pi )\Big]
+q_\Phi\frac{\ln(\Lambda/\mu) }{64\pi^2 R^2}
\Big[ \delta^{\prime\prime}
(y)-\delta^{\prime\prime}
(y-\pi )\Big]
\, .
\label{fiHt}
\end{equation}
Only 
Eq.~(\ref{fiob}) and the first term of 
Eqs.~(\ref{fiH}) and (\ref{fiHt}) (the quadratic divergent part)
contribute to the renormalization of the operator Eq.~(\ref{fit}).
Nevertheless, these contributions   
do not seem to have the advocated form Eq.~(\ref{fit}).
This is because we have not yet demanded that the theory is anomaly
free.
Demanding that the effective 4d theory is free of gravitational
anomalies requires 
$\sum_i q_i=0$, where the sum is over the massless spectrum.
In 5d this implies 
\begin{equation}
\sum_i q_{Q_i}+\sum_i q_{\Phi_i^{++}}=0\, .
\label{anomaly}
\end{equation}
The condition Eq.~(\ref{anomaly})  leads to a total FI-term
(sum of Eq.~(\ref{fiob})  and the first term of Eqs.~(\ref{fiH}) 
and (\ref{fiHt}))
of the form Eq.~(\ref{fit}).
Notice that there is no restriction on the charges of the hypermultiplet
$\Phi^{+-}$ and $\Phi^{c\, -+}$, since they do not contain massless states
(and therefore they  do not contribute to  the 4d gravitational anomaly).
The absence of 4d anomalies does not  guarantee, in general,
a 5d anomaly-free theory \cite{anomalies,fi2}.
Nevertheless the anomalies of the 5d theory can always be cancelled if a 
proper Chern-Simons  term is added \cite{anomalies,fi2}.

The second term of Eqs.~(\ref{fiH}) 
and (\ref{fiHt}) that diverges logarithmically and involve
two derivatives with respect to the extra dimension
renormalizes the operator
\begin{equation}
\int d^4\theta
\,
\frac{\hat\xi_i}{|T|^2}
\partial_5 \Big[\partial_5 V-\frac{1}{\sqrt{2}}
\left(\chi+\chi^\dagger\right)
\Big]\delta(y-y^*_i)
\, ,
\label{fiol}
\end{equation}
where $y^*_i=0,\pi$.
Eq.~(\ref{fiol}) is a brane operator subleading with respect to
Eq.~(\ref{fit})
in an expansion in derivatives $\partial_5^2/\Lambda^2$.
We will
  only be interested in the phenomenological implications
of Eq.~(\ref{fit}) since it  is more sensitive to the cutoff scale.

\section{Phenomenological implications of the FI-terms}

What are the implications of 
the existence of the FI-term  Eq.~(\ref{fit})?
Contrary to the 4d case, the presence of this term 
does not necessary lead to either supersymmetry or gauge symmetry breaking
\cite{agw}.
This is because
the field equation of the auxiliary field  $D$ is  given by
\begin{equation}
D =  -\frac{\partial_{5} \Sigma}{R^2} - \frac{g^2_5}{R} \xi
\Big[\delta(y)-\delta(y-\pi)\Big]\, ,
\label{D}
\end{equation}
and the singlet $\Sigma$ 
can adjust to  guarantee $D=0$ (a supersymmetric vacuum):
\begin{equation}
\langle \Sigma \rangle 
=  -g^2_5 R\xi \int dy\Big[\delta(y)-\delta(y-\pi)\Big]
=-\frac{g^2_5R}{2}\xi\epsilon(y)\, .
\label{sigmavev}
\end{equation}
The  FI-terms 
induce a step-function profile for $\Sigma$.
Since $\Sigma$ couples to the charged hypermultiplets,
these fields get a 5d mass  given by
$
q_\Phi\langle \Sigma \rangle/2
$
that, written in superfields,
 corresponds to a $T$-dependent mass:
\begin{equation}
-\int d^2\mspace{-2mu}\theta\: 
M \, \epsilon(y)\, T\Phi^c\Phi+{\rm h.c.}\ ,\ \ \ \ 
M=\frac{q_\Phi g^2_5}{4}\xi\, .
\label{m5d}
\end{equation}
This is the net result of the FI-terms Eq.~(\ref{fit}) in 5d orbifolds.
From Eq.~(\ref{radion}) we see  that $T$ contains the fifth-component
of the graviphoton $B_5$. This implies that a hypermultiplet
with an induced 5d mass is coupled to the graviphoton. 
This is expected since a 5d mass 
is possible in 5d supergravity
if the   hypermultiplet is charged
under the
graviphoton.
Therefore the effect of the FI-term can also be understood
as a mixing between the graviphoton $B_M$ and the U(1)-photon $A_M$
[$M=(\mu,5)$].
This can be  explicitly seen in Eq.~(\ref{GaugeAb}) 
where a $\langle\Sigma\rangle\not=0$
leads to a mixing between the kinetic term of $A_5$ and $B_5$.
Going to the canonical basis corresponds
to redefine  $A_5$ that in superfields means
$\chi\rightarrow \chi-(\sqrt{2}M\epsilon(y)/q_\Phi) T$.
One can check that by  performing this redefinition
in Eq.~(\ref{GaugeAb}),   the FI-term Eq.~(\ref{fio})
can be eliminated from the theory.
The mixing between $A_\mu$ and $B_\mu$ 
comes from the induced operator
\begin{equation}
\int d^2\mspace{-2mu}\theta\:\xi\epsilon(y)\, 
TW^\alpha_B W_\alpha+{\rm h.c.}\, ,
\end{equation}
where $W^\alpha_B$
is the superfield strength  of the graviphoton $B_\mu$ (up to some normalization
factor).
This operator contains a Chern-Simons  term   
$\varepsilon_{5\mu\nu\rho\sigma}B^5F_B^{\mu\nu} F_A^{\rho\sigma}$
that  is known  not to be  renormalized beyond one-loop.
One can then expect  that the FI-term will  be only renormalized at
the one-loop level \cite{fi2}.

To analyze the implications of the FI-terms, we must 
study the effects of adding a 5d odd mass to the hypermultiplets.
The equations of motion of the scalar fields of the hypermultiplet
are given by  
\begin{eqnarray}
&&\Bigl(\frac{1}{R^{2}} \partial_{5}^{2} +\partial_\mu^{2} - M^{2} - 2\frac{M}{R}
\bigl[\delta(y) - \delta(y-\pi)\bigr]\Bigr) \phi =0\, ,\\
&&\Bigl(\frac{1}{R^{2}} \partial_{5}^{2} + \partial_\mu^{2} - M^{2} + 2\frac{M}{R}
\bigl[\delta(y) - \delta(y-\pi)\bigr]\Bigr) \phi^c =0\, .
\end{eqnarray}
To obtain the 4d  mass spectrum, we can calculate their propagators 
and look at their poles.
These are given  in the Appendix.
For  $\phi^{++}$, we obtain a massless state of wave-function 
\begin{equation}
f(y)=\frac{1}{N_0}e^{M R y}\, ,
\label{wf}
\end{equation}
where $N_0$ is a normalization constant.
For negative (positive)  $M$,
this mode is localized  toward the boundary at $y=0$ ($y=\pi$).
The Kaluza-Klein (KK) spectrum have masses  
$m^{2}_{n} =  M^{2}+n^2/R^2$.
This mass spectrum is paired  up
with that of $\phi^{c\, --}$.

For the hypermultiplet $\phi^{+-}$, we do not have massless states.
The KK states have masses 
$m^{2}_{n} = q_{n}^{2} + M^{2}$, where $q_n$ is given by
$M \tan(q_{n} \pi R) = -q_{n}$.
For $-M\pi R > 1$, 
there is an imaginary  solution for $q_n$
that, 
in the limit $-M R\pi \gg 1$, leads to a KK  mass that is  exponentially
suppressed
\begin{equation} 
m^2 \simeq 4M^2 e^{2M\pi R}\, .
\label{exp}
\end{equation}
The full mass spectrum is paired up 
with that  of $\phi^{c\, -+}$.
From Eq.~(\ref{exp}) we see an interesting effect.
Large 5d odd masses can lead to new light states in the 4d theory.

\section{Implications in theories with Scherk-Schwarz 
supersymmetry breaking} 

In this section we want to analyze the implications
of a 5d odd mass in theories of Scherk-Schwarz (SS) supersymmetry breaking
\cite{ss}.
The SS 
mechanism in superfields
consists in breaking supersymmetry by
turning on a nonzero $F$-term for the radion $T$ 
in  Eqs.~(\ref{GaugeAb}),
(\ref{eq:hyperS}) and (\ref{m5d}).
This gives new mass terms to the 
 hypermultiplet scalars.
To derive the mass spectrum,
it is  more convenient to absorb
these mass terms by redefining the scalar fields (but not the fermion fields):
\begin{equation}
\left(
\begin{matrix}
\phi_{SS}\cr
\phi^{c\, *}_{SS}\cr
\end{matrix}
\right)
=e^{-iF_T\sigma_2 y/2}
\left(\begin{matrix}
\phi\cr
\phi^{c\, *}\cr
\end{matrix}\right)\, .
\label{bcs}
\end{equation}
We will consider the particular case $F_T=1$ (this is 
the value determined dynamically \cite{gqr}).
The boundary terms are not in general invariant under 
the redefinition Eq.~(\ref{bcs}).
For example at $y=\pi$,
where  Eq.~(\ref{bcs}) gives
\begin{equation}
\phi_{SS}(\pi)=-\phi^{c\, *}(\pi )
\ ,\ \ \ \ \ \ \ \ \ 
\phi^{c\, *}_{SS}(\pi )=\phi(\pi )\, ,
\label{bcpir}
\end{equation}
we have that a boundary mass term becomes 
\begin{equation}
M\delta(y-\pi)(|\phi|^2-|\phi^c|^2)
\ \rightarrow \ 
M\delta(y-\pi)(|\phi^c_{SS}|^2-|\phi_{SS}|^2)\, .
\end{equation}
The scalar equations of motion are therefore given by
\begin{eqnarray}
&&\Bigl(\frac{1}{R^{2}} \partial_{5}^{2} +\partial_\mu^{2} - M^{2} - 2\frac{M}{R}
\bigl[\delta(y) + \delta(y-\pi)\bigr]\Bigr) \phi_{SS} =0\, ,\\
&&\Bigl(\frac{1}{R^{2}} \partial_{5}^{2} + \partial_\mu^{2} - M^{2} + 2\frac{M}{R}
\bigl[\delta(y) + \delta(y-\pi)\bigr]\Bigr) \phi^c_{SS} =0\, .
\end{eqnarray}
Eq.~(\ref{bcpir}) also tell us that 
$\phi_{SS}$ must have at $y=\pi$ the 
boundary condition 
   of that of
$\phi^{c\, *}$, and  $\phi^c_{SS}$ 
the boundary condition at $y=\pi$
  of   that of 
$\phi^{ *}$.
For the hypermultiplet of type (a),
this means that $\phi_{SS}$  ($\phi^c_{SS}$)
is  odd (even)
with respect to the boundary at $y=\pi$.
We will denote them as $\phi^{+-}_{SS}$  and 
$\phi^{c\, -+}_{SS}$. 
For the hypermultiplet (b),
$\phi_{SS}$  ($\phi^c_{SS}$)
is even (odd)
with respect to the boundary at $y=\pi$.
We will denote them as 
$\phi^{++}_{SS}$  and 
$\phi^{c\, --}_{SS}$.
 
The scalar $\phi^{++}_{SS}$ 
has a mass spectrum given
by the poles of its propagator calculated in the Appendix.
There is no massless state. For $|M|R\pi\ll 1$ we have
the 4d mass-squared spectrum
\begin{equation}
\frac{2M}{\pi R},\ \frac{1}{R^2},\ \frac{2^2}{R^2}, ...,
\label{small}
\end{equation}
and for $-MR\pi\gg 1$
\begin{equation}
4M^2 e^{MR\pi},\ 
M^2+\frac{1}{R^2},\ 
M^2+\frac{2^2}{R^2}, ...\, .
\end{equation}
Therefore in these two limits, we have a  light scalar (of mass smaller than
$1/R$).
The spectrum of the other scalars, $\phi^{--}_{SS}$, 
$\phi^{-+}_{SS}$ and $\phi^{+-}_{SS}$
is given in the Appendix.

\subsection{A new Scherk-Schwarz  model of electroweak symmetry breaking}

Models of SS supersymmetry breaking
 have been subject of recent research \cite{ab,pq,bhn}. 
Nevertheless,
they  present some phenomenological difficulties, either because
they predict a Higgs too light \cite{pq}
or a compactification scale too low \cite{bhn}.

The presence of  FI-terms (that, as we said, results in 
5d odd masses for the hypermultiplets)
allows to construct new models of
SS supersymmetry breaking.
 Here we will present a new 5d supersymmetric model
similar to those of Refs.~\cite{pq,bhn}
but with the advantage of being able to accommodate
a larger $R^{-1}$ and a Higgs boson of mass close to
the experimental bound.
\begin{table}
\centering   
\begin{tabular}{|c|c|c|}   
\hline   
5d field & Boundary at $y=0$ & Boundary at $y=\pi$ \\   
\hline   
$A_\mu$, $q_3$, $\phi$ & $+$ &$+$ \\   
$\Sigma$, $A_5$, $q^c_3$, $\phi^c$ &$-$&$-$ \\   
$\lambda_1$, $\widetilde Q_3$, $\widetilde\phi$
 & $+$&$-$ \\   
$\lambda_2$, $\widetilde Q_3^c$, $\widetilde\phi^c$ & $-$&$+$ \\   
\hline   
\end{tabular}   
\caption{: \ Boundary conditions, even ($+$) or odd ($-$),
for the bulk superfields.}   
\label{masas}   
\end{table} 
The model consists in a 5d supersymmetric extension of the SM.
The extra dimension will be compactified in a $S^1/\Zparity$
orbifold. The gauge bosons will be living in the 5d bulk, belonging
to a 5d vector supermultiplet. As in Ref.~\cite{bhn}, the Higgs
boson superfields, $\Phi=\{\phi,\widetilde\phi\}$
and $\Phi^c=\{\phi^c,\widetilde\phi^c\}$,
will belong to a 5d hypermultiplet
of  type (b) with zero 5d mass. 
We will discuss later the implication of a  5d mass. 
For simplicity, we will only consider the third family
of quarks. 
The left-handed quark superfield $Q_3=\{q_3,\widetilde Q_3 \}$ 
will be assumed to live in the bulk in a hypermultiplet of type (a) and
 5d  mass $M_3$. 
The right-handed quark superfields, $U_3$ and $D_3$, 
will be respectively localized
 on the  boundaries at $y=0$ and $y=\pi$ \footnote{They could also
 be considered to be living in the bulk with  
large 5d supersymmetric masses such
 that, as we said above,
localize the zero mode toward the corresponding boundaries.}.
This allows the Yukawa couplings
\begin{equation}
{\cal L}_Y=\int d^2 \mspace{-2mu}\theta\: \Big[ Y_t\, \Phi Q_3 U_3
\delta(y)+ Y_b\, \Phi^c Q_3 D_3 \delta(y-\pi)\Big]+{\rm h.c.}\, .
\label{yukawa}
\end{equation}
We will break supersymmetry by the SS mechanism. This means,
as said above, a
change of boundary conditions for the  hypermultiplet scalars
according to Eq.~(\ref{bcpir}).  
We present in Table~1
the resulting  boundary conditions of the bulk fields.
The Higgs hypermultiplet 
will have a massless scalar, $H$, that we will
associate to the SM Higgs.
Using Eq.~(\ref{wf}) and (\ref{yukawa}), we can deduce the top and
bottom mass ratio
\begin{equation}
\frac{m_t}{m_b}=
\frac{Y_t f_{q_3}(0)f_{H}(0)} 
{Y_b f_{q_3}(\pi)f_{H}(\pi)}  
=\frac{Y_t}{Y_b} e^{-M_3\pi R}\, ,
\label{massratio}
\end{equation}
where $f_{q_3}$ is the wave-function of the zero-mode left-handed quark
(given
by Eq.~(\ref{wf})) 
and $f_H=1/\sqrt{\pi R}$ is the wave-function of $H$. 
From the experimental value 
 $m_t/m_b\simeq 60$, we  determine $M_3\simeq -1.3/R$
for $Y_t\simeq Y_b$.
Therefore the effect of the 5d mass $M_3$ is 
to localize the left-handed quark toward the 
boundary at $y=0$ and,  as a consequence,
give  a large
$m_t/m_b$ ratio.
Fixed $M_3$, the model has only one parameter $R$.
All the mass spectrum is determined as a function
of $R$.
Also the SM Higgs potential 
\begin{equation}
V(H)=m_H^2 H^2
+\lambda H^4\, ,
\end{equation}
is determined as a function of $R$.
Let us calculate it.
Contributions to $m_H^2$ arise at the loop level
from  gauge and top interactions.
They can be calculated following Ref.~\cite{pq}. We obtain
\begin{equation}
\Delta m_H^2= \left[3 g_2^2+ \frac{3}{5} g_1^2 \right]
  \Pi_V        + 6 h^2_t\,  \Pi_{Q_3}\, ,
\label{oneloop}
\end{equation}
where $g_i$  and $h_t$ are the 4d gauge  and top couplings and
\begin{eqnarray}
\label{bulkcorr}
       \Pi_V& =&\pi R\int \frac{d^4 p}{(2\pi)^4}\,
       \left[G_V^{++}
            -G_V^{+-}\right]\, , \\
       \Pi_{Q_3}& =&\frac{1}{f^2_{q_3}(0)}\int \frac{d^4 p}{(2\pi)^4}\,
       \left[G_{Q_3}^{+-} 
            -G_{Q_3}^{++}\right]\, , 
\end{eqnarray}
and the propagators $G_V$ and $G_{Q_3}$ are given in the Appendix.
In the limit $-M_3\gg 1/(\pi R)$,
 the top contribution becomes small.
This is because  in this limit
the KK states of $Q_3$
become infinitely massive and only   a massless
chiral  superfield,  localized on the  boundary,
remains in the theory.
For our value $M_3\simeq -1.3/R$ we are close to this regime.
Hence the quark sector of our  model is similar
to that  of Ref.~\cite{pq} 
where the quarks are  localized on the boundary.
It is known that in this case one-loop corrections
give masses to the squark on the boundary
\begin{eqnarray}
m_{\widetilde Q_3}^2&=& \left[\frac{16}{3} g_3^2+ 
3 g^2_2+\frac{1}{15} g_1^2 \right] 
\Pi_V+2h^2_t\Pi_V\, ,\\
m_{\widetilde U_3}^2&=& \left[\frac{16}{3} g_3^2+\frac{16}{15} g_1^2 \right] 
\Pi_V+4h^2_t\Pi_V\, ,
\end{eqnarray}
and, at the two-loop level, a sizeable contribution  to $m_H^2$ \cite{pq}:
\begin{equation}
\Delta m^2_H\simeq \frac{3h^2_t}{16\pi^2 }
(m^2_{\widetilde Q_3}+m^2_{\widetilde U_3})\ln[R^2m^2_{\widetilde
Q_3}]\, .
\label{twoloop}
\end{equation}
Up to now we have assumed that the 5d mass of the Higgs hypermultiplet
is zero. Nevertheless, we already saw that it can be generated by
radiative corrections due to a FI-term of the U(1)$_Y$.
From Eqs.~(\ref{m5d})
and (\ref{small}) we see that a 5d mass for the Higgs hypermultiplet
implies a mass for $H$.
This is given by
\begin{equation}
\Delta m^2_H\simeq \frac{2M}{\pi R}
= \frac{g^2_1Y_H}{2}\xi
\simeq \frac{ g^2_1 Y_H}{32\pi^2}\Lambda^2\, ,
\label{5dmasscon}
\end{equation}
where $Y_H$ is the Higgs hypercharge.
This contribution depends on the ultraviolet
cutoff $\Lambda$ and therefore cannot be predicted.
For $\Lambda\sim 1/R$ this is around  a $25\%$ of the others.
This tell us that, if this contribution is present, 
we will have large uncertainties in
our calculation of the electroweak scale. 
Nevertheless, we must say that since the presence of a 5d mass $M$ for the Higgs
is related to the anomalies of the theory, it could
be possible that the contribution Eq.~(\ref{5dmasscon})  is zero, 
as it happens in certain models \cite{pq}.

Finally, the quartic coupling in $V(H)$ 
is fixed by supersymmetry to be
 $\lambda=(g_2^2+3g^2_1/5)/8$. 
It also receives sizeable loop corrections 
that in our case, where  $M_3\simeq -1.3/R$,
 are approximately given by
\begin{equation}
\Delta \lambda=\frac{3 h_t^4}{16\pi^2}\ln
\left(\frac{m^2_{\widetilde Q_3}+m^2_t}{m_t^2}\right)\, .
\label{oneloopl}
\end{equation}
Combining all the contributions to the Higgs potential, we can study
the breaking of the electroweak symmetry.
We see that 
the negative contribution to $m^2_H$,
Eq.~(\ref{twoloop}), dominates over the positive gauge contribution
triggering  the electroweak symmetry breaking.
Fixing the vacuum expectation value  of $H$ to be 174 GeV
and neglecting Eq.~(\ref{5dmasscon}), 
we obtain a value of $R^{-1}= 3-4$ TeV.
This large value of the compactification scale $1/R$ 
guarantees small corrections (from the KK) 
to the electroweak observables.
This prediction, however, turns to be very sensitive to the value
of $m_t$ and $g_3$ which  have large
experimental uncertainties. 
This is because 
the positive and negative contributions to 
$m^2_H$ are very close in size, cancelling each other for some values
of the top mass inside the experimental window.
This accidental cancellation  does not allow us  to get a precise 
determination of $R$ (even when Eq.~(\ref{5dmasscon}) is neglected).
Nevertheless, the physical Higgs mass is not very much sensitive to $R$ 
and can be determined with more precision.
We obtain $m_{Higgs}\simeq 110$ GeV,
with a theoretical uncertainty that we estimate of  $5-10$ GeV.
We see then that the Higgs is around the present experimental bound.
Therefore this model can be confirmed or excluded
in the near future.

\section{Conclusions}

We have studied the effect  of FI-terms
in 5d orbifold theories.
These terms can be induced at the one-loop level
by boundary or bulk fields.
Their form  is dictated 
by the gauge symmetry and supersymmetry
and can be easily derived using
$N=1$ superfields. By introducing
the
radion superfield $T$ we were able to
 understand the effect of the FI-term as a mixing between the graviphoton and 
the U(1) gauge boson. 
In superfields this corresponds to a mixing between $\chi$ and $T$,
and between
$W^\alpha$ and $W^\alpha_B$.
This allows us to relate the FI-term to the 
Chern-Simons  term  
graviphoton-graviphoton-photon \cite{fi2}.
The net effect of the FI-term is  to induce 5d odd masses 
for the hypermultiplets.

We  have then analyzed how the 4d mass spectrum is modified if 
hypermultiplets have  5d  odd masses.
For large values of these masses, we have seen that
  new light states can be present in the effective 4d theory.
We have also considered the modifications in the mass spectrum
of theories 
with  Scherk-Schwarz supersymmetry breaking.
The existence of 
5d odd masses allows for variations on previous models \cite{pq,bhn}.
We have presented a new model where the left-handed quark is in
the bulk and has a 5d mass $M_3$. The  value of $M_3$ can be fixed to explain
the $m_t/m_b$ ratio.
The model has only one parameter $R$ and it is a priori very predictive.
We find however 
accidental cancellations in the  contributions to the Higgs potential
that makes difficult to
determine  the electroweak scale. 
Higher loop contributions must be considered.
Nevertheless, the massive spectrum is completely determined as a function of $R$.
The massless spectrum corresponds to the one of the SM.
The model also predicts a Higgs  boson with a mass
around the present experimental bound.

{\bf Acknowledgments:}

We thank Riccardo Barbieri, Guido Marandella and Michele Papucci
for sharing with us  their recent work related with 
the one presented here. 
This work was partially supported by the CICYT Research Project
AEN99-0766 and DURSI Research Project 2001-SGR-00188.

\newpage

\section*{Appendix. Propagators of the 5d fields}

In this Appendix we will derive the propagators of the
hypermultiplet scalars, $\phi$ and $\phi^c$, with a
5d odd mass  $M\epsilon(y) R$.
Some of them have also been  derived in Ref.~\cite{fi2}.

\paragraph{ 1) Supersymmetric case:}

The propagators in the mixed position-momentum space are defined by
\begin{align}
    \Bigl(\frac{1}{R^{2}} \partial_{5}^{2} - p^{2} - M^{2} -
    2\frac{M}{R}
    \bigl[\delta(y) - \delta(y-\pi)\bigr]\Bigr) G_{p}^{\phi}(y,y') &=
   -\frac{1}{R} 
\delta(y-y') \label{eq:Greenphi}\:,\\
    \Bigl(\frac{1}{R^{2}} \partial_{5}^{2} - p^{2} - M^{2} +
    2\frac{M}{R} \bigl[\delta(y) - \delta(y-\pi)\bigr]\Bigr)
    G_{p}^{\phi^{c}}(y,y') &= -   \frac{1}{R} \delta(y-y')
    \label{eq:Greenphic} \:,
\end{align}
where $p$ is the  Euclidean 4d momentum. Below we will show the propagators
for the scalar $\phi$  for different boundary conditions.
The propagators for $\phi^c$ can be obtained from those of $\phi$
just by replacing $M\rightarrow -M$.
 
\paragraph{Even-Even:} 
For the case (a) defined in section~2 
the scalar  $\phi$ is even at both boundaries,
and therefore the two source terms in \eqref{eq:Greenphi} will
affect the propagator behaviour through boundary conditions. The
propagator is given by
\begin{multline}
    G^{\phi^{++}}_{p}(y,y') = \frac{\cosh \bigl[p_M R y_{<}\bigr] +
      \frac{M}{p_M} \sinh \bigl[p_M R y_{<}\bigr]}{p_M\Bigl([1 -
      M^2/p_M^2]
      \sinh \bigl[p_M \pi R\bigr]\Bigr)} \\
    \times \biggl(\cosh \bigl[p_M R (\pi - y_{>})\bigr] -
    \frac{M}{p_M} \sinh \bigl[p_M R (\pi - y_{>})\bigr]\biggr) \,,
\end{multline}
with $p_M \equiv \sqrt{p^{2} + M^{2}}$ and where $y_{<}$ ($y_{>}$) denote
the lesser (greater) of $y$ and $y'$.
The pole at $p_M=M$ corresponds to the massless mode, with a 
wave-function given by
\begin{equation}
    f(y)=\bigl[\Res_{p_M^2=M^2}
    G_{p}(y,y)\bigr]^{1/2} = \frac{1}{\sqrt{(e^{2MR\pi}-1)/(2MR)}} e^{MR
    y}\:, 
\label{eq:zeromdwf}
\end{equation}
The poles at $p_M=\im n/R, n=1, 2,\ldots$
lead to a spectrum of massive modes given by 
$m^{2} = - p^{2} = M^{2} - p_M^{2}=
(n/R)^{2} + M^{2}$.

\paragraph{Odd-Odd:}
In this case the propagator must vanish
at both boundaries. We have
\begin{equation}
    G^{\phi^{--}}_{p}(y,y') =  \frac{\sinh \bigl[p_M
      y_{<}\bigr]\sinh \bigl[p_M R
      (\pi - y_{>})\bigr]}{p_M \sinh \bigl[p_M \pi R\bigr]}\,.
\end{equation}
The  mass spectrum  is the same as  the massive sector  of
the even-even case.

\paragraph{Even-Odd:}
In the  case (b), since $\phi$ (and $G_p^{\phi^{+-}}$) vanishes at the
$y=\pi$ boundary it cannot feel the effect of the source term at
$y=\pi$. Solving \eqref{eq:Greenphi} with these boundary conditions gives rise to
\begin{equation}
    G^{\phi^{+-}}_{p}(y,y') =  \frac{\Bigl(\cosh \bigl[p_M R y_{<}\bigr] +
      \frac{M}{p_M} \sinh \bigl[p_M R y_{<}\bigr]\Bigr)\sinh \bigl[p_M R
      (\pi - y_{>})\bigr]}{p_M\Bigl( \cosh \bigl[p_M \pi R\bigr] +
      \frac{M}{p_M} \sinh \bigl[p_M \pi R\bigr]\Bigr) }\,.
\label{propeo}
\end{equation}
There are non-trivial poles located at the imaginary axis $p_M=\im q_{n}$,
where $q_{n}$ are the solutions of $M \tan(q_{n} \pi R) = -q_{n}$. The
mass spectrum is then given by $m^{2}_{n} = q_{n}^{2} + M^{2}$.
When $ - M\pi R > 1$ there is also a  pole at the real axis.
In the limit $- M\pi R \gg 1$ this corresponds to a state 
with  an  exponentially-suppressed mass, $m^2 \simeq 4M^2 e^{2M\pi R}$.

\paragraph{Odd-Even:}
We have 
\begin{equation}
    G^{\phi^{-+}}_{p}(y,y') =   \frac{\sinh \bigl[p_M R y_{<}\bigr]
      \Bigl(\cosh \bigl[p_M R (\pi - y_{>})\bigr] -\frac{M}{p_M} 
      \sinh \bigl[p_M R (\pi - y_{>})\bigr]\Bigr)}{p_M\Bigl(
      \cosh \bigl[p_M \pi R\bigr] -\frac{M}{p_M} \sinh \bigl[p_M \pi R\bigr]\Bigr)}\,,
\end{equation}
For $\phi^{c\, -+}$ we must replace $M\rightarrow -M$ which leads
to the same 4d mass spectrum as that of $\phi^{+-}$.

\paragraph{2) Scherk-Schwarz  supersymmetry breaking:}

The boundary mass terms at $y=\pi$ in the equations of motion change
as a result of the SS mechanism as explained above.
We have now
\begin{align}
    \Bigl(\frac{1}{R^{2}} \partial_{5}^{2} -
    p^{2} - M^{2} - 2\frac{M}{R}
    \bigl[\delta(y) + \delta(y-\pi)\bigr]\Bigr) G_p^{\phi_{SS}} 
(y,y^\prime)&=
       -\frac{1}{R}  \delta(y-y') \:,\\
    \Bigl(\frac{1}{R^{2}} \partial_{5}^{2} -
    p^{2} - M^{2} + 2\frac{M}{R}
    \bigl[\delta(y) + \delta(y-\pi)\bigr]\Bigr) 
G_p^{\phi^{c}_{SS}}(y,y^\prime) &=    -\frac{1}{R} \delta(y-y') \:.
\end{align}
The only propagators that are modified with respect to the 
supersymmetric case are  those which are even at $y=\pi$. 

\paragraph{Even-Even:}
\begin{multline}
    G^{\phi^{++}_{SS}}_{p}(y,y') =  \frac{\cosh \bigl[p_M R y_{<}\bigr] +
      \frac{M}{p_M} \sinh \bigl[p_M R y_{<}\bigr]}{p_M\Bigl( \bigl[1 + (M^2/p_M^2)
      \bigr]
      \sinh \bigl[p_M \pi R\bigr] + 2 \frac{M}{p_M}\cosh \bigl[p_M\pi
      R\bigr] \Bigr)}\\
    \times \biggl(\cosh \bigl[p_M R (\pi - y_{>})\bigr] +
      \frac{M}{p_M} \sinh \bigl[p_M R (\pi -
      y_{>})\bigr]\biggr) \, .
\end{multline}
There are    poles at the real axis in the
$p_M$ complex plane, given by the solutions of the equation
\begin{equation}
    \label{eq:tanhnos}
    \tanh (p_M \pi R) = -\frac{2p_M/M}{1 +
    (p_M/M)^{2}}\,.
\end{equation}
For $|M| R\pi \ll 1$ we get a scalar of mass-squared $m^{2} \simeq 2M/(\pi R)$,
while for $-MR\pi \gg 1$ there are two light scalars with mass-squared
given by $m^2\simeq \pm 4M^2 e^{M\pi R}$.  There are also poles at the
imaginary axis, placed at $p_M = \im q_{n}$, where $q_{n}$ is the
solution to the equation $\tan (p_M \pi) = - (2 M/p_M)/[1 -
(M/p_M)^{2}]$.  For $M R\pi\ll 1$ the KK mass spectrum is $m_n^2\sim n^2/R^2$.
\paragraph{Odd-Even:}
\begin{equation}
    G^{\phi^{-+}_{SS}}_{p}(y,y') = \frac{\sinh \bigl[p_M Ry_{<}\bigr]
      \Bigl(\cosh \bigl[p_MR(\pi  - y_{>})\bigr]+ \frac{M}{p_M}
      \sinh \bigl[p_MR (\pi  - y_{>})\bigr]\Bigr)}{p_M\Bigl(
      \cosh \bigl[p_M \pi R\bigr] + \frac{M}{p_M} \sinh \bigl[p_M \pi
      R\bigr]\Bigr)}\,.
\end{equation}
The pole structure is the same as that in Eq.~(\ref{propeo}).

From the above  propagators we can   derive  the propagators 
of the vector and quark sector  that are used in section~4.1.
They are simply  given by $G_V^{++}=G_p^{\phi^{++}}$ and 
$G_V^{+-}=G_p^{\phi^{+-}}$  in the limit $M=0$.
For  $Q_3$, we have $G_{Q_3}^{++}=G_p^{\phi^{++}}$ 
and $G_{Q_3}^{+-}=G_p^{\phi^{+-}}$ 
with $M=M_3$.

\end{document}